\documentclass[9pt,twocolumn,twoside]{osajnl}

\usepackage[squaren, Gray, cdot]{SIunits}

\journal{ao}

\setboolean{shortarticle}{false}

\ifthenelse{\boolean{shortarticle}}{\colorlet{color2}{color2b}}{\colorlet{color2}{color2}} 

\title{Proof of the feasibility of a nanocell-based wide-range optical magnetometer}

\author[1,2*]{Emmanuel Klinger}
\author[1]{Hrayr Azizbekyan}
\author[1]{Armen Sargsyan}
\author[2]{Claude Leroy}
\author[1]{David Sarkisyan}
\author[1]{Aram Papoyan}

\affil[1]{Institute for Physical Research -- NAS of Armenia, Ashtarak-2, 0203 Armenia}
\affil[2]{Laboratoire Interdisciplinaire Carnot de Bourgogne -- UMR CNRS 6303, Universit{\'e} Bourgogne  Franche-Comt{\'e}, BP 47870, 21078 Dijon Cedex, France}

\affil[*]{Corresponding author: emmanuel.klinger@u-bourgogne.fr}

\dates{Compiled \today}

\ociscodes{(280.4788) Optical sensing and sensors; (230.3810) Magneto-optic systems; (300.6210) Spectroscopy, atomic; (020.7490) Zeeman effect; (020.2930) Hyperfine structure.}

\doi{\url{http://dx.doi.org/10.1364/josab.XX.XXXXXX}}

\begin{abstract}
We present an experimental scheme performing scalar magnetometry based on the fitting of Rb D$_2$ line spectra recorded by derivative selective reflection spectroscopy from an optical nanometric-thick cell. To demonstrate its efficiency, the magnetometer is used to measure the inhomogeneous magnetic field produced by a permanent neodimuim-iron-boron alloy ring magnet at different distances. The computational tasks are realized by relatively cheap electronic components: an  Arduino Due board for the external control of the laser and acquisition of spectra, and a Raspberry Pi computer for the fitting. The coefficient of variation of the measurements remains under $5\%$ in the magnetic field range of 40 -- 200 mT, limited only by the size of the oven and translation stage used in our experiment. The proposed scheme is expected to operate with a high measurement precision also for stronger magnetic fields ($>500~$mT), in the hyperfine Paschen-Back regime, where the evolution of the atomic transitions can be calculated with a high accuracy.
\end{abstract}

\setboolean{displaycopyright}{true}

\begin{document}

\maketitle
\thispagestyle{fancy}

\ifthenelse{\boolean{shortarticle}}{\ifthenelse{\boolean{singlecolumn}}{\abscontentformatted}{\abscontent}}{}

\section{Introduction}
From the past decades, laser spectroscopy of atoms has established as the core of high precision measurements \cite{demtroder}, finding a large variety of applications ranging from testing of the fundamental symmetries of nature to the detection of magnetic field from the heart and the brain. Resonant interaction of laser radiation with atomic vapor have been used for development of optical insulators \cite{weller_ol_2012}, narrow-band optical filters \cite{kiefer_sr_2014}, stabilization of laser radiation frequency \cite{su_ao_2014}, determination of fundamental constants \cite{truong_nc_2015}, etc. Among the important applications emerging from atomic spectroscopy is atom-based sensing, which is now used in metrology (atomic clocks) \cite{ludlow_rmp_2015},  nuclear magnetic resonance gyroscopes,  interferometers \cite{kitching_ieee_2011}, and optical magnetometers \cite{budker_book_2013}.
Most of atomic sensors do not require cryogenic cooling, which is advantageous for integration in measurement systems and their miniaturization \cite{yang_nph_2007}.

Atomic spectroscopy of thermal alkali metal vapors exposed to magnetic fields underlying the optical magnetometry has been intensively studied and explored in the past decades (for the reviews, see \cite{budker_book_2013, arimondo_rmp_1977, budker_rmp_2002}). Modern state-of-the-art magnetometers allow to achieve remarkable sensitivity \cite{budker_np_2007}, and are mostly focused on the measurement of extremely low fields in shielded environment. They are successively used for measurements of biomagnetic fields \cite{biso01,taue01}, revealing hidden ferromagnetic objects \cite{shei01}, etc. But besides, there are many other applications such as measurement and mapping of high gradient magnetic field in nuclear magnetic tomography, remote monitoring of nuclear reactors, alignment of particle accelerators, etc., where high spatial resolution, immunity against external perturbations, large dynamic range of measurement, and robust, autonomous, unshielded operation are of key priority, rather than unprecedented sensitivity attained by implementing sophisticated and expensive measurement schemes.

While Hall gauges are the most employed sensors to perform magnetometry of fields ranging from tens to thousands of Gauss, they are not suitable for performing remote sensing and measurement in strongly perturbing radiation environment such as in nuclear reactors \cite{bolshakova_2007}. This task can be solved using optical magnetometers, which are immune against electric perturbations and thermal drift. 
Earlier it was demonstrated that optical nanocells (NC) having a gap between the windows of 40 -- 1000$~$nm are very convenient to perform sub-Doppler spectroscopy \cite{carteleva_josab_2009}, and can serve as a promising tool for sensing applications. As it has been shown in recent works, implementation of the technique of selective reflection (SR) from NC, and notably recording its derivative spectra (dSR), allows one to reduce the resonance linewidth down to $\sim40$~MHz \cite{sargsyan_jetpl_2016}, which is strongly beneficial when studying spectra of alkali metal vapors placed in magnetic fields \cite{klinger_epjd_2017, sargsyan_jpb_2018}.

In this article, we present the proof-of-concept for an optical magnetometer based on spectroscopy of Rb vapor contained in a NC. A performance analysis of the developed laboratory prototype magnetometer is presented, realized by exposing the NC to the inhomogeneous magnetic field produced by a permanent neodymium-iron-boron alloy ring magnet at different distances.
The article is structured as follows. In Section \ref{sec:setup}, we present the optical and electronic scheme of the magnetometer, as well as the measurement procedure and corresponding algorithms. Performance analysis of the magnetometer is presented in Section \ref{sec:charact}, addressing precision, response time, and performance limitations.

\section{Measurement technique}
\label{sec:setup}

The employed measurement technique combines optical, electronic and programming components. Here we give a step by step description of the experimental assembly, methodology, and measurement algorithms. It is worth noting that the proposed system can only measure magnetic field component directed along the laser beam.

\begin{figure}[h!]
\centerline{\includegraphics[scale =0.7]{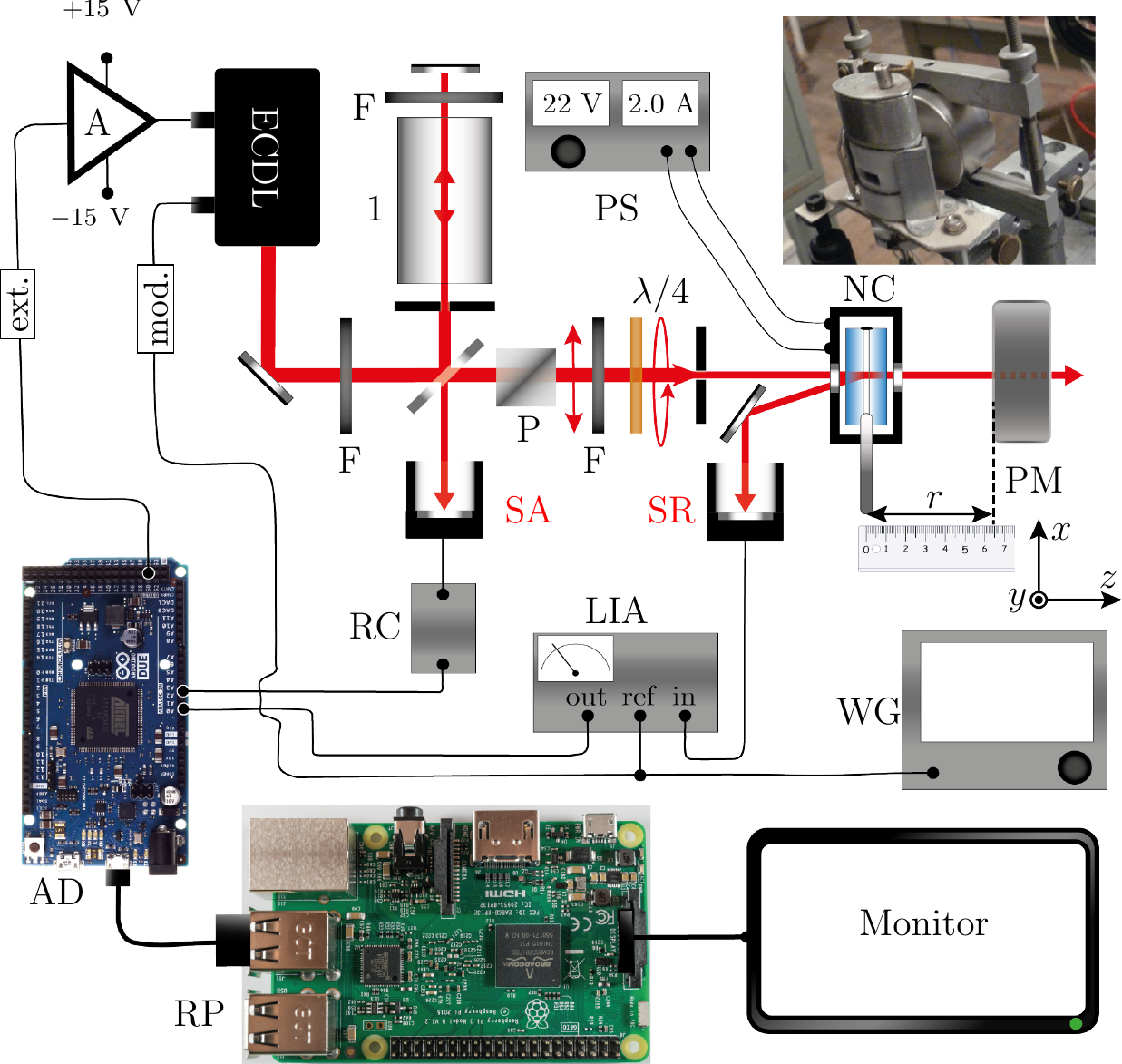}}
\caption{Experimental arrangement of the magnetometer. ECDL -- extended cavity diode laser; F -- filters; 1 -- reference Rb cell; P -- polarizer; $\lambda/4$ -- quarter-wave-plate; PS -- DC power supply; NC -- nanocell + oven assembly; PM -- ring magnet; WG -- waveform generator; LIA -- lock-in amplifier; RC -- RC circuit; A -- amplification circuit; AD -- Arduino Due board; RP -- Raspberry Pi computer; SA and SR -- photodetectors recording saturated absorption and selective reflection signals. Upper-right inset: photograph of the nanocell in the oven and the magnet.}
\label{fig:experimentalSetup}
\end{figure}

\subsection{Experimental arrangement}

\subsubsection{Optical elements}
The measurement of magnetic fields was carried out on the setup presented in Fig.~\ref{fig:experimentalSetup}. An externally controlled extended cavity diode laser (ECDL) emits a linearly-polarized light ($\lambda_L=780~$nm, $\Delta\omega_L\simeq 2\pi\times 1~$MHz) tuned to the Rb D$_2$ line. A fraction of the light was sent to an auxiliary branch to form a saturated absorption frequency reference, recorded with a photodetector (SA) equipped with a cylindrical guiding pipe to prevent contribution from residual ambient light to the detected signal.
The polarization of the light in the main branch was purified by a polarizer (P) and then converted to a left-hand circularly-polarized light using a quarter-wave-plate ($\lambda/4$) in order to excite $\sigma^+$ transitions in the Rb vapor. The beam was directed at normal incidence onto the NC. The arising selective reflection beam was carefully separated from other backward beams and sent to the second photodetector (SR) using a mirror.

In order to reach sufficient atomic density of 2.0$\times$10$^{13}~$cm$^{-3}$, the sidearm of the cell was heated to $125~^\circ$C using a 3 cm-thick oven, while keeping the temperature of the cell windows at least $25~^\circ$C higher to prevent condensation of vapor on the sapphire windows. A set of filters was used to adjust the power of the laser light to $\sim 1~$mW for the SA branch, and $\sim \unit{100}{\micro\watt}$ for the beam directed onto the NC. Pinholes were used to keep the incident beam diameter small enough ($\sim 1~$mm) for covering a uniform thickness area of the cell. 

A neodymium-iron-boron alloy ring magnet (PM) with an axis aligned along the laser beam was placed at the rear side of the NC-oven assembly to apply a longitudinal magnetic field. The magnet was mounted on a translation stage, and the magnet--NC distance $r$ was measured using a simple ruler. The strength of the magnetic field applied to the vapor column was changed by a simple longitudinal translation of the magnet. We have carefully checked that the beam was completely transmitted through the center of a 6~mm-diameter hole of the ring magnet for all the positions of the magnet, which enables to minimize the residual transverse magnetic field components.

\subsubsection{Electronics}
The linear scanning of the laser radiation frequency across the Rb D$_2$ line was controlled by an Arduino Due board (AD), which outputs a 0 to +3.3~V voltage with a 12 bit resolution ($0.80~$mV steps), followed by a self-made amplification circuit (A), which was used for signal conditioning to extend the voltage range to -7 to +7~V needed for the external ramping control of ECDL. In addition to scanning, the laser frequency was also modulated by a 5 kHz sine wave of 300 mV amplitude generated by a Siglent SDG 5082 waveform generator (WG). This modulation signal served as a reference for the SRS SR510 lock-in amplifier (LIA), which picked the signal from SR photodetector and directly generated the dSR signal on the output. The time constant of the LIA was set to the minimum to prevent broadening distortion of the dSR spectrum.

Both the SA and SR signals were processed by an analog-to-digital converter of the AD board, and sent to the Raspberry Pi (RP) computer via USB port. The output voltage of the SA photodetector was kept between 0 and +3.3~V to meet the AD range. This was done by the use of an RC circuit serving simultaneously for attenuation of the output voltage of SA photodetector, and for filtering the high-frequency components arising because of the sine modulation of the laser frequency. The output voltage of the lock-in amplifier was also carefully adjusted to fit 0 to +3.3~V range, which was realized by applying offset voltage (dSR signal can be slightly negative) and choosing an appropriate sensitivity level.

\subsection{Measurement procedure}
\label{sec:Mathematicaalgo}

The procedure of measurement can be summarized by the following actions: initiation, acquisition, and fitting. The interaction between the user and the magnetometer relies on a Mathematica-based program processed on the RP computer and displayed on a monitor. A summary of the main program is shown in the Algorithm$~$\ref{alg:Mathematica} table. This simple program is based on four different functional blocks that we have developed in Mathematica, and which we detail hereafter.

\begin{algorithm}[h]
\caption{Mathematica algorithm performing the measurement}
\label{alg:Mathematica}
\begin{algorithmic}[1]
\small\Procedure{MeasureMagneticField}{}
\State $\textcolor{magenta}{data}$ = Initiation();
\State $\textcolor{magenta}{recordedSpectrum}$ = getSpectrum();
\State $\textcolor{magenta}{rescaledSpectrum}$ = rescaleFrequency($\textcolor{magenta}{recordedSpectrum}$);
\State $\textcolor{magenta}{magneticFieldValue}$ = fitMagneticField($\textcolor{magenta}{data}, \textcolor{magenta}{rescaledSpectrum}$);
\State \textbf{return} $\textcolor{magenta}{magneticFieldValue}$
\EndProcedure
\end{algorithmic}
\end{algorithm}

\subsubsection{Initiation}
The initiation function can be summarized as follows. First, four data files containing the transition frequency shifts and transition amplitudes versus magnetic field (two for $^{85}$Rb, and two for $^{87}$Rb) are imported. We have limited the magnetic field span to a realistic range of 10 -- 200~mT conditioned by particular magnet, oven, and translation stage used in our experiment. The step of magnetic field variation is set to 0.5~mT, such that the number of lines per file (590) does not overload the RAM of the RP computer. Furthermore, an additional file is being loaded to interpolate the lineshape of the dSR signal versus the laser frequency, see Sub-section \ref{subsec:fitMagneticField} for more details.

\subsubsection{getSpectrum}
Upon the user demand, the RP computer commands a single-shot scanning of the laser frequency across the Rb D$_2$ line to AD, which is executed by generating a sweep voltage. For each value of the generated sweep voltage, AD reads the analog input signal to save the output voltage from the SA photodetector and the dSR signal from the LIA. 

After each step of voltage sweeping generated by the AD digital-to-analog converter, the AD makes a 60~ms pause before reading the SA and dSR signals in order to let the atom-light interaction regime to be established. This procedure ensures obtaining as narrow spectra as possible, taking into account that a too long interaction time leads to optical pumping, which is not accounted in the fitting procedure. At the end of the sweep, the saved values are retrieved by the RP.  

\subsubsection{rescaleFrequency}
The re-scaling step is necessary in order to get a common $x$-axis for the experimental and the theoretical spectra (as the variable is voltage for the first case, and frequency in the second). For that the program requests data from the SA channel: using the FindPeaks routine of Mathematica, the program identifies the strongest crossover resonances of the group of transitions $^{85}$Rb $F_g = 3 \rightarrow F_e = 2,3,4$ and $^{85}$Rb $F_g = 2 \rightarrow F_e = 1,2,3$.

The zeroing of the frequency axis is also needed in order to fit correctly the experimental signal, and for that we thus choose the strongest crossover resonance of $^{85}$Rb $F_g = 2 \rightarrow F_e = 1,2,3$, located at $\omega = 2\pi\times 384.232125082$~THz from the ground state $5~^2S_{1/2}$ \cite{steck_85Rb}, as the zero detuning.

\subsubsection{fitMagneticField}
\label{subsec:fitMagneticField}

The fitting procedure relies on the theoretical analysis presented in  \cite{klinger_jcp_2018}, where a modeling of the optical interaction of the laser with Rb atoms in a NC in the presence of magnetic field was presented. As one can see from Eq.~(19) in \cite{klinger_jcp_2018}, the accounting of the atomic velocity distribution in the NC for calculating SR spectrum can be extremely time consuming, as it is not possible to get analytic solutions. However, assuming that all the transitions in the spectrum have the same lineshape (which in the case of a NC is neither Loretzian/Gaussian nor Voigt), the selective reflection signal (Eq.~(14) in \cite{klinger_jcp_2018}) can be re-written as:
\begin{equation}
S_r(\omega,B) = \sum_j \mu_j(B) \times S(\omega-\omega_j(B)),
\end{equation}
likewise the dSR signal can simply be expressed as 
\begin{equation}
dS_r(\omega,B)\equiv \frac{\partial }{\partial \omega} S_r(\omega,B) = \sum_j \mu_j(B) \times \frac{\partial }{\partial \omega} S(\omega-\omega_j(B)),
\end{equation}
where $\mu_j(B)$ and  $\omega_j(B)$ are part of the data obtained with the initiation function. The summation on $j$ is done for each transition present in the frequency window returned by rescaleFrequency. The lineshape $S(\omega,\ell, \Gamma)$ of each individual transitions was pre-calculated using successively Eqs. (19), (18) and (14) from \cite{klinger_jcp_2018}, setting $\mu_{eg}=1$ and $\Delta=0$, while the cell thickness $\ell=490\pm10~$nm and the linewidth $\Gamma=50\pm5~$MHz were fit to that of a recorded experimental Rb D$_2$ spectrum (without magnetic field). This simplifications lead to a large enhancement of the calculations speed and, in this way, we estimate the calculation time to be reduced by a factor 100.

Using the acquired dSR spectrum with a re-scaled $x$-axis and the data from the initiation, the program recursively calculates theoretical spectra (natural Rb: $72\%$ of $^{85}$Rb; $28\%$ of $^{87}$Rb) in the given frequency window, by changing the magnetic field with a step of 0.5~mT. For every calculated spectra, the program looks for a qualitative agreement between experiment and theory by making differences of the signals for all scanned frequencies, and calculates their average. The returned magnetic field value by the routine is the one having the smallest averaged residuals. To get a slightly better precision than 0.5~mT on the returned value, we made use of the interpolation function of Mathematica on the $B$-field dependence of the averaged residuals.

\section{Performance analysis}
\label{sec:charact}

\begin{figure*}[h!]
\centerline{\includegraphics[scale = 1]{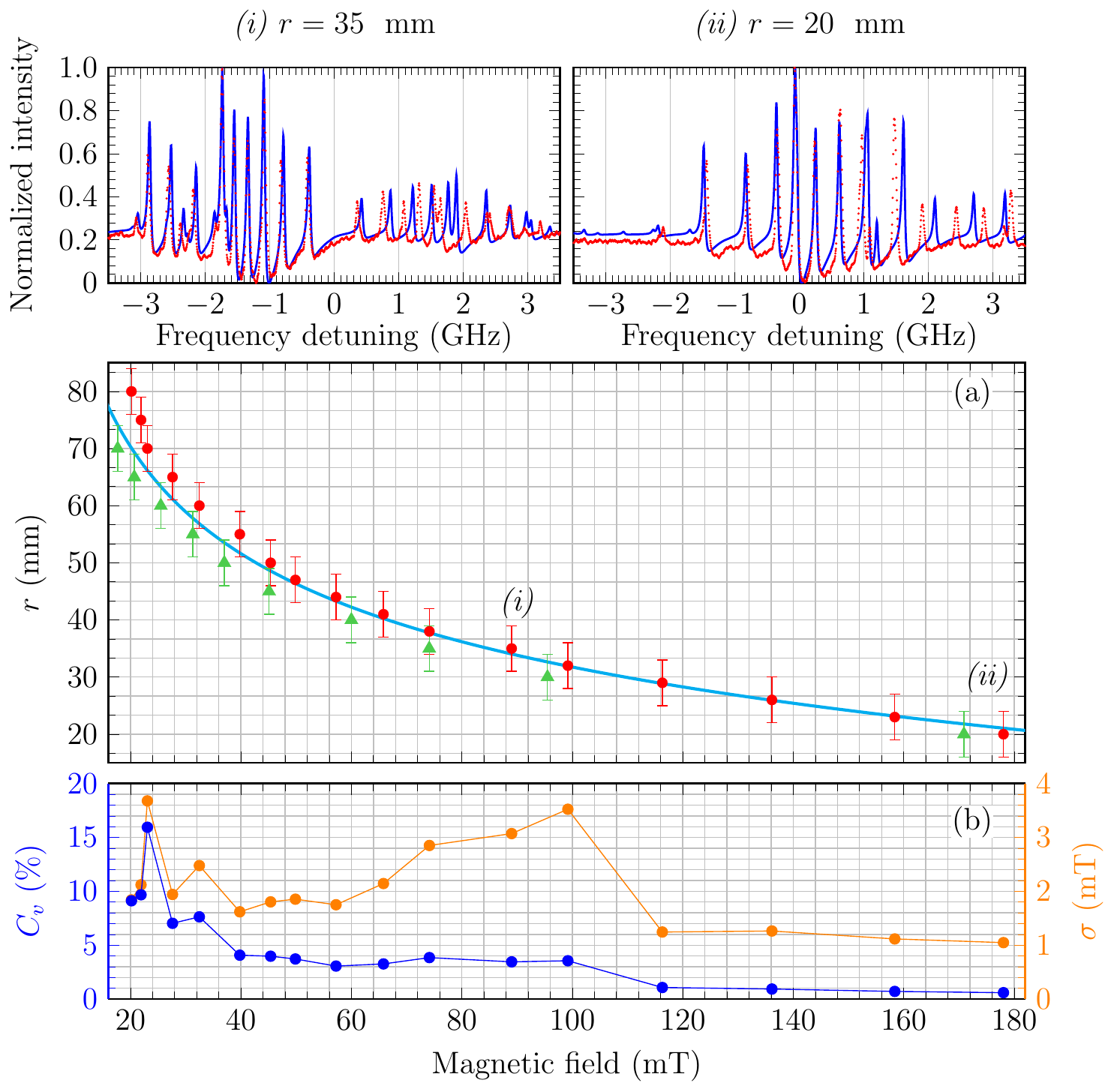}}
\caption{(a) Measured magnetic field ($x$ axis) as a function of the cell-magnet distance $r$ ($y$ axis) averaged over 20 measurements. Red dots: experimentally measured with NC; green triangles: measured with a Hall gauge; solid line: calculated. (b) Evolution of the standard deviation $\sigma$ (orange line, right $y$-axis), and the coefficient of variation $C_v$ (blue line, left $y$-axis) as a function of the measured magnetic field. The insets $(i,ii)$ show two obtained spectra (red dots) and their respective fits (blue solid lines) recorded for $r=35$, 20~mm and corresponding to the measured value of $B= 89.5$ and 179~mT, respectively. Both the experimental and theoretical spectra are plotted with an increment $n=2$, see the text.}
\label{fig:calibratedMagnet}
\end{figure*}

The following methodology was implemented in order to test the magnetometer and evaluate its performance. The distance $r$ between the ring magnet and the NC was varied from 20 mm to 80 mm with a 3 mm step using translation stage, preserving co-axial alignment of the magnet in respect to the laser beam. The distance was monitored with a 1~mm graduated ruler ($\Delta r \simeq 4~$mm). For each value of $r$, the measurement procedure described in the previous section was employed, returning the measured $B$ value. The results of these measurements showing the response of our setup to the longitudinal magnetic field are presented in Fig.~\ref{fig:calibratedMagnet}.  

The examples of spectra recorded and processed during the measurement procedure are presented in the insets $(i)$, $(ii)$ of the figure, corresponding to $r = 35$, and 20~mm, respectively. 
The traces of red dots represent  experimentally recorded spectra, and the best fitting simulations are shown by blue lines. Based on such fitting runs performed for 17 values of $r$, we have built a $B$-field calibration curve (see Fig.~\ref{fig:calibratedMagnet}a). Red dots on this graph indicate the $B$-field averaged over 20 values obtained by our measurement procedure, the green triangles correspond to a calibration curve obtained with a 'Teslameter HT201' magnetometer, and the solid line is the calculated dependence following \cite{peng_jmmm_2004}
\begin{equation}
\begin{aligned}
B(r) = \frac{B_{res}}{2}\Bigg[& \frac{r}{\sqrt{r^2 + R_{out}^2}}-\frac{r-h}{\sqrt{r^2-h^2 + R_{out}^2}} \\
&\qquad- \frac{r}{\sqrt{r^2 + R_{in}^2}}+\frac{r-h}{\sqrt{r^2-h^2 + R_{in}^2}}\Bigg],
\end{aligned}
\end{equation}
where $R_{in} = 3$~mm and $R_{out} = 25$~mm are the inner and outer radius respectively, $h=30$~mm is the height and the producer-specified residual induction $B_{res}$ at the magnet surface of 1.45~T. An obvious consistency of the measured magnetic field with the calibration curve of the magnet justify applicability of the proposed concept. 

\subsection{Precision}
To estimate the precision and repeatability of the measurement technique, we have calculated the standard deviation $\sigma$, and the coefficient of variation  $C_v = \sigma/\mu$ where $\mu$ is the mean of $B$, determined for each positions of the magnet from the results of twenty consecutively recorded and fitted spectra (Fig.~\ref{fig:calibratedMagnet}b). For measured $B$-fields over $40~$mT, we get $C_v < 5\%$, which is a good value for a non-optimized table-top laboratory setup. Higher precision is expected for stronger magnetic field, as the establishment of a hyperfine Paschen-Back regime leads to equalization of the amplitudes of individual transitions and increases their frequency shift \cite{sargsyan_jpb_2018}. As expected, the deviation increases with the decrease of the magnetic field, caused by transition overlapping.
But even in this case,  the relative deviation still remains below $10\%$. 

\subsection{Response time}
Another important characteristic of the magnetometer is its temporal resolution. In the case of the proposed measurement technique, the overall response time is the time needed to return the measurement result. After measuring the response time of each routine, it appeared that the recording and rescaling are taking about 30~s, while the fitting routine itself lasts tens of minutes.

To accelerate the temporal response of the magnetometer, we have checked the behavior of the fitting routine when increasing the increment $n$ for the laser frequency sampling. For $n=1$, the measurement readout is taken from all the 4096 frequency points separated by $\approx 2.2~$MHz. For $n=2$, the readout is taken at every 2nd point (2048 points in total), etc. Increasing the increment allows to reduce the overall number of operations, thus reducing the response time in a close to linear way for small values of $n$: $\tau (n) \approx \tau (n=1)/n$. For $n=1$, the measured response time was about 20 minutes. The measurement process can be accelerated by increasing the sampling interval. But the increase of the $n$ value eventually leads to distortion of the recorded spectral lineshape, thus reducing the overall measurement precision. Our studies have shown that no difference in returned results (recorded spectra) is observed for $n\leqslant 18$ in the explored range of magnetic field; non-negligible differences appear for $n>20$. At the same time, the above mentioned initial nearly-linear reduction of measurement time with the increase of sampling interval saturates for $n\geqslant10$ on the level of $\tau \lesssim 2~$minutes.

\subsection{Limitations}
Precision and sensitivity of the proposed magnetometer are limited by random and systematic errors. The main sources of random errors are: i) electrical noises in measurement circuits; ii) mechanical vibrations; iii) thermal convection of air in the vicinity of NC windows. Among the most impacting sources of systematic errors are: i) non-linearity in laser frequency scanning; ii) non-perfect circular polarization of the input laser radiation; iii) non-uniformity of the NC thickness across the incident laser beam; iv) imperfect normal incidence angle and/or large divergence of the laser beam. Less contribution is expected from variation of the NC temperature. Nevertheless, we should note that most of these errors can be strongly reduced in a final design of a magnetometer device.

As most table-top systems, our magnetometer suffers from the environmental perturbances. Shielded wiring, integration of electronics in a chip device and connection matching can strongly suppress electric noise. The impact of mechanical (acoustical) vibrations causing laser beam walk across the tapered gap of the NC, resulting in  change of the cell thickness during the scanning and impacting the signal lineshape, can be strongly reduced in the case of a compact design and low-height rigid mounting of optical components. Moreover, this impact can be reduced using a NC with a large homogeneous thickness, or a NC with pressure-controlled thickness \cite{sargsyan_jetp_2017}.

Similarly, solutions can be found to overcome influence of systematic effects. Non-linearity of the laser frequency scanning, which is obviously seen in the upper insets of Fig.~\ref{fig:calibratedMagnet}, is  caused by imperfect grating control of the ECDL and can be avoided with the use of a free-running laser diode, which will also be a cost-effective solution. However, in this case the measurement precision can be lowered for small magnetic fields due to the 10 -- 20 MHz linewidth of such lasers. Alternatively, a Fabry-Pérot etalon could be used to perform the frequency re-scaling. Proper polarization control can be easily reached by using commercially available good-quality polarizer and quarter-wave plate. As to the non-uniformity of the cell thickness in the region of laser beam and improper incidence angle or beam divergence, these problems will be mostly solved when implementing a proper design of the compact device, as was mentioned above. Although rising the NC side arm temperature over $130~^\circ$C could improve the signal-to-noise ratio, it also leads to additional spectral broadening of the atomic lines, which mostly impacts the lower $B$-field measurement limit ($B<30~$mT), when most of individual lines overlap. Finally, the possible light intensity fluctuations can also be suppressed in a cheap way by implementing a feedback control, as is shown in \cite{truong_pra_2012}.

\section{Conclusion \& Outlook}
To conclude, we have shown that NCs filled with Rb vapors are a good tool to perform scalar magnetometry. We have presented a user-friendly experimental scheme realizing the measurement with coefficient of variation $< 5\%$ in the range 40 -- 200~mT which should be unchanged for magnetic fields of several hundreds of mT.  

Let us note that optical vapor magnetometry in the Voigt configuration using a mm-long cell was reported in \cite{keaveney_jpb_2019} using Stokes polarimetry \cite{weller_jpb_2012}. Carrying out similar measurements using NC is a natural outlook of this work. Indeed, it should be possible, using the same scheme as presented in this article, to measure the magnetic field applied in the plane of polarization of the laser beam. Besides, extending our fitting procedure to the second circularly polarized component should allow revealing also the sign of the measured $B$-field: for left-hand circularly polarized incident light, $\sigma^+$ ($\sigma^-$) excitation occurs for positive (negative) projection of $B$-field along the light propagation axis correspondingly, and vice-versa for the right-hand circularly polarized incident light.

Such magnetometer can be applied for numerous physical problems and beyond. Indeed, one of the advantages of NC-magnetometry is the possibility to measure high-gradient fields with a sub-$\mu$m resolution thanks to the ultra-small thickness of the vapor column. Having a proper theoretical model, it will be also possible to measure static electric fields (Stark effect) using the same procedure. In addition, it is worth mentioning that the selective reflection signal can be reliably detected at a far distance (tens of meters) from the NC thanks to a low beam divergence, which also proves that this magnetometer is a good candidate for remote sensing magnetic fields, for instance in the difficult environments faced in high energy physics, space and civil engineering. 

Miniaturization of spectroscopic kits for metrological applications has recently received much attention \cite{hummons_optica_2018}. Aiming at development of a compact low-cost commercial magnetometer device, we have used in our work relatively cheap electronic components. We should note that also optical nanocells are becoming more affordable as new techniques for their manufacturing are being developed \cite{peyrot_ol_2019}.  

\section*{Funding Information}
E.K. acknowledges a travel grant for collaborative research from FAST (Foundation for Armenian Science and Technology).

\end{document}